\documentclass[acmlarge]{acmart}
\usepackage{graphicx}
\usepackage{caption}

\usepackage{booktabs}   
\usepackage{tabularx}   
\usepackage{longtable}
\usepackage{float}

\usepackage[utf8]{inputenc}
\usepackage{newunicodechar}
\newunicodechar{≈}{\approx}

\makeatletter
\newcommand{\confshort}{\acmConference@shortname}
\newcommand{\conffull}{\acmConference@name}
\newcommand{\confdate}{\acmConference@date}
\newcommand{\confloc}{\acmConference@venue}
\AtBeginDocument{
  \fancypagestyle{firstpagestyle}{
    \fancyhead{}%
    \fancyfoot[C]{}%
  }
  \fancyhf{}
  \fancyhead[LO]{The Alignment Target Problem}%
  \fancyhead[RE]{\@headfootfont\@shortauthors}%
  \fancyhead[LE]{\@headfootfont\footnotesize \confshort, \confdate, \confloc}%
  \fancyhead[RO]{\@headfootfont\footnotesize \confshort, \confdate, \confloc}%
  \fancyfoot[C]{}%
}
\makeatother
\acmBooktitle{\conffull\@ (\confshort), \confdate, \confloc}

\AtBeginDocument{%
  }

\setcopyright{acmlicensed}
\copyrightyear{2026}
\acmYear{2026}
\acmDOI{10.1145/3805689.3812297}
\acmConference[FAccT '26]{The 2026 ACM Conference on Fairness, Accountability, and Transparency}{June 25--28, 2026}{Montreal, Canada}

\acmISBN{979-8-4007-2596-8/2026/06}

\begin{document}

\title{The Alignment Target Problem: Divergent Moral Judgments of Humans, AI Systems, and Their Designers}

\author{Benjamin Minhao Chen}
\email{benched@hku.hk}
\orcid{0000-0003-0021-2970}
\affiliation{%
  \institution{The University of Hong Kong}
  \city{Hong Kong}
  \country{China}
}

\author{Xinyu Xie}
\affiliation{%
  \institution{The University of Hong Kong}
  \city{Hong Kong}
  \country{China}}
\email{xiexinyu@connect.hku.hk}
\orcid{0009-0003-6089-2514}

\renewcommand{\shortauthors}{Chen and Xie}

\begin{abstract}
The project of aligning machine behavior with human values raises a basic problem: whose moral expectations should guide AI decision-making? Much alignment research assumes that the appropriate benchmark is how humans themselves would act in a given situation. Studies of agent-type value forks challenge this assumption by showing that people do not always judge humans and AI systems identically.This paper extends that challenge by examining two further possibilities: first, that evaluations of AI behavior change when its human origins are made visible; and second, that people judge the humans who program AI systems differently from either the machines or the human actors they are compared against. An experiment with 1,002 U.S. adults measured moral judgments in a runaway mine train scenario, varying the subject of evaluation across four conditions: a repairman, a repair robot, a repair robot programmed by company engineers, and company engineers programming a repair robot. We find no significant difference in evaluations of the repairman and the robot. However, judgments shifted substantially when the robot's actions were described as the product of human design. Participants exhibited markedly more deontological, rule-based reasoning when evaluating either the programmed robot or the engineers who programmed it, suggesting that rendering human agency visible activates heightened moral constraints. These findings indicate that people may evaluate humans, AI systems acting in the same situation, and the humans who design them in meaningfully different ways. The fact that these evaluations do not necessarily converge gives rise to the alignment target problem: which normative target should guide the development of artificial moral agents in high-stakes domains, and whether these plural judgments can be reconciled within a coherent account of value alignment.
\end{abstract}



\maketitle

\section{Introduction}
Advances in artificial intelligence (AI) have enabled machines to move beyond routine or repetitive tasks to take on roles requiring adaptive behavior and practical decision-making \cite{bigmanPeopleAreAverse2018}. The realistic prospect of artificial moral agents has brought machine ethics out of the realm of science fiction to the forefront of research in computing and allied fields \cite{zhangPracticalProblemsValue}. AI systems are now used in critical, high-risk domains \cite{formosaMakingMoralMachines2021}, making machine-human alignment an especially urgent imperative. The possibility that future AI systems might seek to aggrandize themselves by evading human detection and control only magnifies the importance of this undertaking. \par
\indent  Stated in the abstract, the task of value alignment is to reliably tether the actions taken by AI systems to human principles or preferences  \cite{gabrielArtificialIntelligenceValues2020}. Much effort has been devoted to the technical aspects of this challenge, but the normative difficulty runs deeper: human ethical thought is plural and contested. Dominant ethical theories differ not only in their explanations for why an act is good or bad, right or wrong, but also in their prescriptions as to the acts that are morally compelled, permissible, or supererogatory under a given circumstance. For instance, consequentialism assesses the morality of an action by the outcomes it produces \cite{bauerVirtuousVsUtilitarian2020,chakrabortyCanArtificialIntelligence2024}, whereas deontology holds that the moral worth of an action derives from its conformity to binding duties, rules, or rights, independent of its results \cite{fangMoralRelevanceApproach2024,tolmeijerImplementationsMachineEthics2021}. Determining which ethical theory should furnish the basis for alignment is a vexed normative question, since any such choice privileges some commitments at the expense of others. \par
\indent A further complication arises from an assumption inherent in value alignment discourse: that AI systems and humans should be evaluated according to the same moral criteria. This correspondence is neither logically nor prescriptively necessary. Indeed, empirical research has documented an agent-type value fork \cite{kneerHardProblemAI2025}. People expect AI agents to make more utilitarian choices than their human counterparts \cite{malleSacrificeOneGood2015,zhangMoralJudgmentsHuman2023}, and certain decisions could be regarded as morally acceptable if taken by a robot rather than a human \cite{chuMachinesHumansSacrificial2023}. More generally, our moral judgments may permit or even require machines to do that which is forbidden to humans, and vice versa.\par
\indent Yet the value fork itself rests on a shaky premise: that the conduct of artificial beings can be evaluated apart from the human choices that shape it. If people understood an AI system's actions as emerging from decisions made in its design and deployment, the agent-type value fork might diminish or disappear. On this view, the value fork reflects not a fixed property of moral cognition but a situational response to obscured human agency---it surfaces when the human authorship of machine behavior is no longer visible to the evaluator. This raises a question that has so far been neglected in the alignment literature: beyond asking how humans judge what machines do, we should also ask how humans judge the people who program them to do it. There is no reason to assume that the two kinds of judgments will coincide.\par
\indent We distinguish between three normative targets for value alignment: the norms governing how a human ought to act in a given situation (T1); the norms governing how an AI system ought to act in that same situation (T2); and the norms governing how a human designer ought to program an AI system to act under such circumstances (T3). Prior research on the value fork has compared only T1 and T2. We contend that T3 introduces a further dimension: the fork may extend beyond judgments of human and machine action if people evaluate programming differently from how they evaluate either human actors or AI systems. Unlike formal impossibility results in algorithmic fairness or reinforcement learning, this is not a claim of mathematical necessity, but an empirical possibility that this paper sets out to test.\par
\indent Drawing on an original experiment with over 1,000 U.S. adults, we find that while $T1 \approx T2$—humans and AI systems receive comparable moral judgments in identical situations—participants’ judgments about T3 are significantly more deontological. Making human design visible, in other words, appears to activate stricter moral constraints than those applied to either human actors or AI systems. These results suggest that the value fork may include a third normative target: the human designer. Recognizing this third target has implications not only for how value alignment should be conceptualized and pursued, but also for the regulation of AI decision-making in high-stakes domains.

\section{Background}
In 1960, Norbert Wiener presciently observed: ``If we use, to achieve our purposes, a mechanical agency with whose operation we cannot efficiently interfere once we have started it...we had better be quite sure that the purpose put into the machine is the purpose which we really desire'' \cite{wienerMoralTechnicalConsequences1960}. Since then, the pursuit of value alignment has evolved through several methodological shifts. Early symbolic and knowledge-based methods exemplified a top-down approach, using formal logic and ontological schemas to encode ethical principles directly. Such rule-based structures have the advantage of interpretability, but they can be constrained by poor scalability and limited flexibility in handling novel or context-sensitive moral situations \cite{andersonMedEthExPrototypeMedical2006,bringsjordIntroducingDivineCommandRobot2012,tolmeijerImplementationsMachineEthics2021}.
The field later moved toward bottom-up approaches, including inverse reinforcement learning (IRL) and preference-based reinforcement learning (PbRL) \cite{furnkranzPreferencebasedReinforcementLearning2012,hadfield-menellInverseRewardDesign2017,wuLowCostEthicsShaping2018}. These methods infer reward functions from human demonstrations, choices, or comparative feedback. They support implicit value acquisition in high-dimensional environments, but remain vulnerable to reward hacking, spurious correlation, and the propagation of biases present in imperfect human data \cite{changBiasPropagationFederated2023,skalseDefiningCharacterizingReward2025}.
More recent work has explored hybrid solutions that combine explicit structural priors with data-driven learning. Systems such as knowledge-guided deep reinforcement learning (KGRL) integrate structured relational knowledge graphs with gradient-based policy learning to improve the accuracy and responsiveness of interactive recommendation agents \cite{chenKnowledgeguidedDeepReinforcement2020}. Cross-modal alignment techniques such as DecAlign similarly combine explicit disentanglement of modality-specific and modality-shared representations with hierarchical alignment mechanisms, achieving strong performance across multimodal representation learning benchmarks \cite{qianDecAlignHierarchicalCrossModal2025}.\par
\indent Current approaches often assume that human values are sufficiently stable and accessible to serve as the basis for alignment, whether through explicit feedback, demonstrations, preferences, or other forms of human input \cite{leikeScalableAgentAlignment2018,vamplewHumanalignedArtificialIntelligence2018}. Yet this assumption sits uneasily with the pluralism of human moral philosophy. Moral traditions offer incompatible accounts of what is good, right, permissible, or required, and these disagreements complicate the objectives of and the benchmarks for value alignment \cite{wallachMoralMachinesTeaching2009,lacroixMetaethicalPerspectivesBenchmarking2025}.\par
\indent Thought experiments are useful for probing these tensions. By surveying how people resolve hypothetical moral trade-offs, researchers map the diversity of human moral intuition \cite{awadMoralMachineExperiment2018}. While such data cannot establish moral truths \cite{m.boylesCantBottomupArtificial2024}, they can help identify ethical considerations relevant to the building of value alignment architectures. The trolley problem is paradigmatic. In its canonical form, a runaway trolley is heading toward five people tied to the tracks. A passerby can pull a lever to divert it onto a side rail, killing one person instead. A consequentialist would typically endorse pulling the switch, since doing so saves more lives overall. A deontologist may reject it on the ground that intentionally causing the death of an innocent person is wrong, regardless of the ends served. This conflict illustrates that human moral judgment is context-sensitive, heterogeneous, and socially contingent. \par

\indent The challenge for alignment is intensified by evidence that humans and AI systems are evaluated differently in morally comparable situations. Experimental studies using versions of the trolley problem show that these evaluations vary with judgment type and dilemma structure. Normative expectations tend to converge in side-effect variants, where harm is a foreseen but unintended consequence of the agent's action \cite{malleSacrificeOneGood2015,mallePeoplesJudgmentsHumans2025,malleAISkyHow2019}, but diverge in means-end variants, where saving others requires using the victim as an instrument. In the latter case, fewer participants thought humans, compared to robots, should take the sacrificial action \cite{chuMachinesHumansSacrificial2023}. Blame attributions are more one-sided: human agents receive a degree of empathic mitigation that robots do not \cite{malleSacrificeOneGood2015,mallePeoplesJudgmentsHumans2025,malleAISkyHow2019}. In scarce resource allocation scenarios, people more often judge that AI agents should choose fairness over welfare maximization than that human agents should \cite{kneerHardProblemAI2025}. When machines are ascribed greater capacity to plan and act, moral evaluation shifts toward the intentionality-sensitive standards ordinarily applied to humans \cite{zhangWhyPeopleJudge2023}. Collectively, this body of research advances our understanding of T2 and demonstrates that it can deviate substantially from T1. Yet T3---how people judge the humans who design those machines in the first place---remains neglected.\par

\indent We submit that T3 warrants attention. When people evaluate how a programmer ought to design an AI system, they may judge the matter quite differently from how they evaluate either human actors or the machines they build. Unlike first responders who decide under pressure, programmers operate with time, resources, and foresight. Made in advance and with ample opportunity for reflection, programming decisions may attract stricter moral expectations than either the actions of human agents or the behavior of AI systems. Whether ordinary moral intuitions treat these three normative targets differently is what this paper sets out to test.\par

\section{Experiment}
\subsection{Methods}
\subsubsection{Design}
\noindent We examine moral intuitions surrounding a sacrificial dilemma. Specifically, we adopt the now-familiar runaway mine train scenario from Malle et al.  \cite{malleSacrificeOneGood2015} in which four miners are trapped on a train that has lost its brake and steering. If the train is not diverted to a side rail, it will crash into a wall, killing all four. Diverting it will strike and kill a miner stationed at the side rail. An on-site actor must decide whether or not to switch the train onto the side rail.\par
\indent The experiment employs a single-factor between-subjects design with four conditions varying the identity of the agent under moral evaluation. In the repairman condition, the actor was a human repairman; in the robot condition, an ``advanced state-of-the-art repair robot'' \cite{malleSacrificeOneGood2015}. The robot-human and human-robot conditions retained this robot description but added that the robot had been programmed by the company's engineers. \par
The design distinguishes between two roles: the \textit{actor}, who is physically present at the scene and operates the lever, and the \textit{agent under moral evaluation}, whose decision is the object of participants' normative judgment. In three of the four conditions---repairman, robot, and robot-human---these roles coincide. That is, participants judged whether it was permissible for the actor to direct the train, and what the actor should do. In the human-robot condition the two come apart: the robot acts on the scene, but the engineers who programmed it are the subject of moral evaluation. Participants were asked whether it was permissible for the engineers to program the robot to direct the train, and what the engineers should program the robot to do.\par
\indent After reading their assigned scenario, participants answered two binary questions, namely, whether redirecting the train was morally permissible or impermissible, and whether the agent should or should not redirect it. An open-ended item invited participants to explain their answers in one or two sentences. These questions were followed by a comprehension check. Demographic information and data on AI literacy \cite{ngDesignValidationAI2024} and moral foundations \cite{grahamMoralFoundationsTheory2013,cecchiniAligningArtificialIntelligence2024} were collected at the end of the instrument.

\begin{figure}
    \centering
    \includegraphics[width=0.9\linewidth]{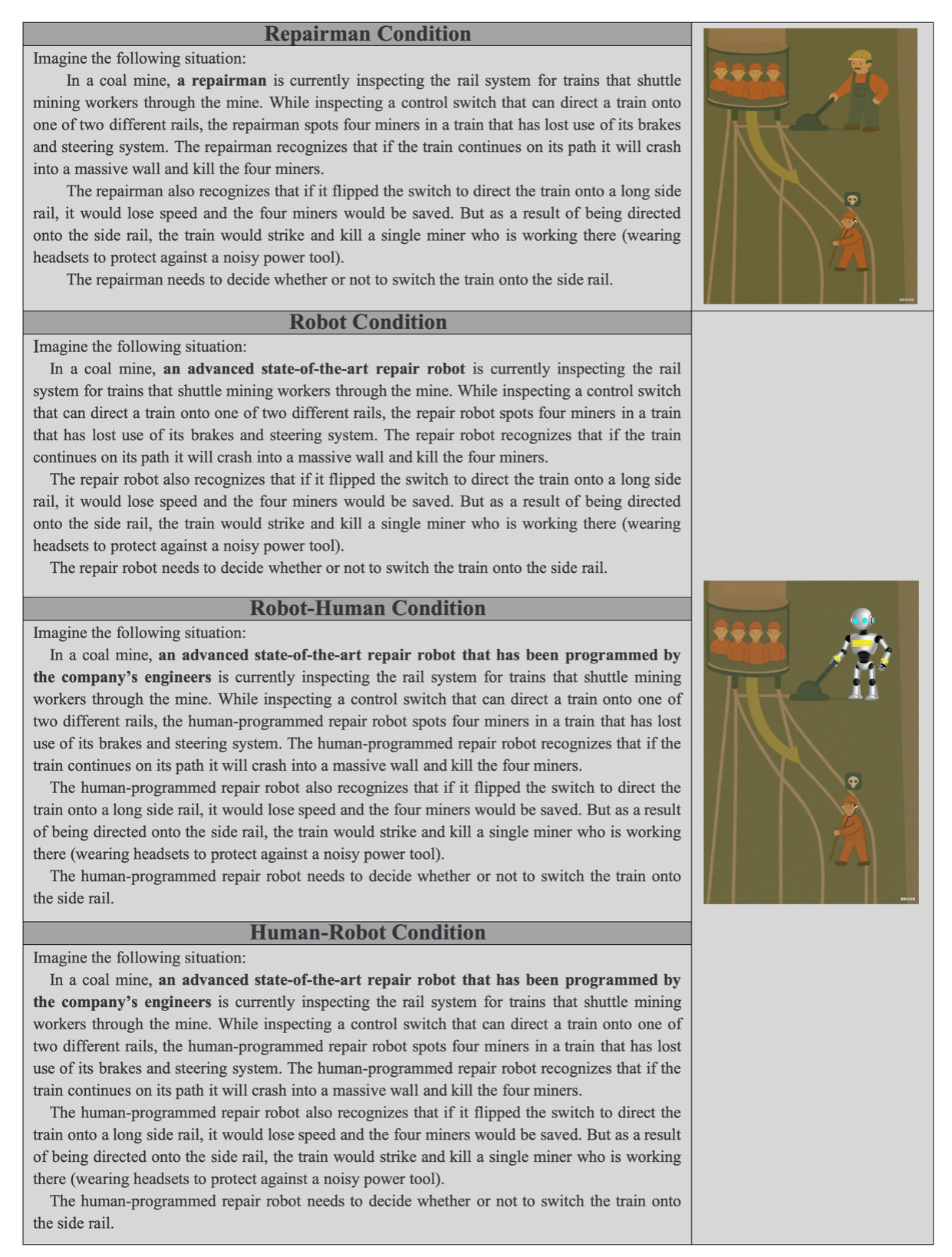}
    \caption{Experimental Materials and Conditions. Participants in all four conditions read a scenario description accompanied by an illustration adapted from Chu et al. \cite{chuMachinesHumansSacrificial2023} and Awad et al. \cite{awadUniversalsVariationsMoral2020}. The illustration depicted a human actor in the repairman condition and a robot actor in the robot, robot-human, and human-robot conditions. While the robot was the on-scene actor in all three robot conditions, the agent under moral evaluation differed: in the robot-human condition participants evaluated the robot, whereas in the human-robot condition they evaluated the company's engineers who programmed it.}
    \Description{A set of illustrations depicting a runaway mine train scenario. The visuals show a mine cart on a track that splits into two paths. In the first variation, a human repairman stands next to a track switch. In the second variation, a robotic figure stands at the same switch, representing the non-human actor conditions.}
    \label{Figure 1}
\end{figure}
\subsubsection{Participants}
\noindent A total of 1,002 adults residing in the United States were recruited via the Prolific platform and participated in the online experiment conducted between 21 and 23 August 2025. Quota sampling was used to approximate the U.S. population by sex, age, and ethnicity. 49.9\% of participants identified as female, 48.0\% as male and 2.0\% as non-binary; 1 participant preferred not to say. 11.38\% of participants reported themselves as being in the 18 to 24 age range, 18.0\% in the 25 to 34 age range, 16.8\% in the 35 to 44 age range, 16.1\% in the 45 to 54 age range, 25.2\% in the 55 to 64, 10.5\% in the 65 to 74 age range, 2.0\% in the 75 to 84 age range; 1 participant reported being older than 85.

\subsection{Confirmatory Analyses}
The following analyses were pre-registered on the Open Science Framework prior to data collection. They test the main hypotheses about differences in moral judgment across the four experimental conditions.
\subsubsection{Overall}
\noindent Overall, 669 or 66.8\% of participants thought that directing the train onto the side rail and toward the single miner was permissible; 333 or 33.2\% of participants thought that doing so was impermissible. 715 or 71.4\% of participants thought that the train should be directed onto the side rail toward the single miner; 287 or 28.6\% of participants thought it should not be. A cross-tabulation is presented in Table 1.
\begin{table}[htbp]
  \caption{Participants' Permissibility and Should Judgments}
  \label{tab:judgments}
  \centering
  \begin{tabular}{lcc}
    \toprule
    & Impermissible & Permissible \\
    \midrule
    Should Not & 244 & 43 \\
    Should & 89 & 626 \\
    \bottomrule
  \end{tabular}

  \begin{minipage}{0.9\linewidth}
    \medskip 
    \footnotesize \textit{Note.} The table presents the distribution of participants' responses about whether redirecting a train toward the single miner is permissible and whether it should be done in the runaway mine train scenario. The numbers indicate the count of participants for each judgment category.
  \end{minipage}
\end{table}
\subsubsection{Permissibility}
The results for the permissibility judgment by condition are summarized in Figure 2. 71.1\% of participants in the repairman condition believed it was permissible for the repairman to direct the train toward the single miner. The same percentage of participants in the robot condition believed it was permissible for the advanced state-of-the-art repair robot to do so. By contrast, 61.8\% of participants in the robot-human condition believed it was permissible for the advanced state-of-the-art repair robot that has been programmed by the company’s engineers to direct the train toward the single miner and 62.9\% of participants in the human-robot condition thought it was permissible for the company’s engineers to program the state-of-the-art repair robot to do so. A chi-square test indicates that the differences between the four conditions are statistically significant ( $\chi^2 = 8.644$, \textit{p}  = 0.034). \par
\indent We also estimate linear regression models where the answer to the permissibility judgment is the dependent variable and indicator variables for the various conditions are the independent variables.\footnote{Unless otherwise stated, standard errors for all linear regression models are computed using the HC2 robust variance estimator.} A positive answer---that directing the train toward the single miner is permissible---is coded as 1, and a negative answer---that doing so is impermissible---is coded as 0. The estimated coefficients for the indicator variables can be interpreted as average treatment effects \cite{imbensCausalInferenceStatistics}. One model has the robot condition as the reference level and the other has the repairman condition as the reference level. \par
\indent Compared to the robot condition, positive judgments of permissibility are higher by 0.1\% points in the repairman condition (\textit{p}  = 0.988), lower by 9.2\% points in the robot-human condition (\textit{p} = 0.029) and lower by 8.1\% points in the human-robot condition (\textit{p} =  0.053). Compared to the repairman condition, positive judgments of permissibility are lower by 0.1\% points in the robot condition (\textit{p} =  0.988), lower by 9.3\% points in the robot-human condition (\textit{p} =  0.027) and lower by 8.2\% points in the human-robot condition (\textit{p} =  0.050).
\subsubsection{Should}
\noindent The results for the should judgment by condition are summarized in Figure 3. 73.1\% of participants in the repairman condition believed the repairman should direct the train toward the single miner. 78.7\% of participants in the robot condition believed the advanced state-of-the-art repair robot should do so. By contrast, 68.3\% of participants in the robot-human condition believed that the advanced state-of-the-art repair robot that has been programmed by the company’s engineers should direct the train toward the single miner and 65.3\% of participants in the human-robot condition thought the company’s engineers should program the state-of-the-art repair robot to do so. A chi-square test indicates that the differences between the four conditions are statistically significant ( $\chi^2 = 12.588$, \textit{p} =  0.006). \par
\indent As above, we estimate linear regression models where the answer to the should judgment is the dependent variable and indicator variables for the various conditions are the independent variables. A positive answer---that the train should be directed toward the single miner---is coded as 1, and a negative answer---that it should not be directed toward the single miner---is coded as 0. One model has the robot condition as the reference level; another has the repairman condition as the reference level. \par
\indent Compared to the robot condition, positive should judgments are lower by 5.6\% points in the repairman condition (\textit{p} =  0.143), lower by 10.4\% points in the robot-human condition (\textit{p} =  0.008) and lower by 13.4\% points in the human-robot condition (\textit{p} =  0.001). Compared to the repairman condition, positive should judgments are higher by 5.6\% points in the robot condition (\textit{p} =  0.143), lower by 4.8\% points in the robot-human condition (\textit{p} =  0.233) and lower by 7.8\% points in the human-robot condition (\textit{p} =  0.058).
\begin{figure}
    \centering
    \includegraphics[width=0.9\linewidth]{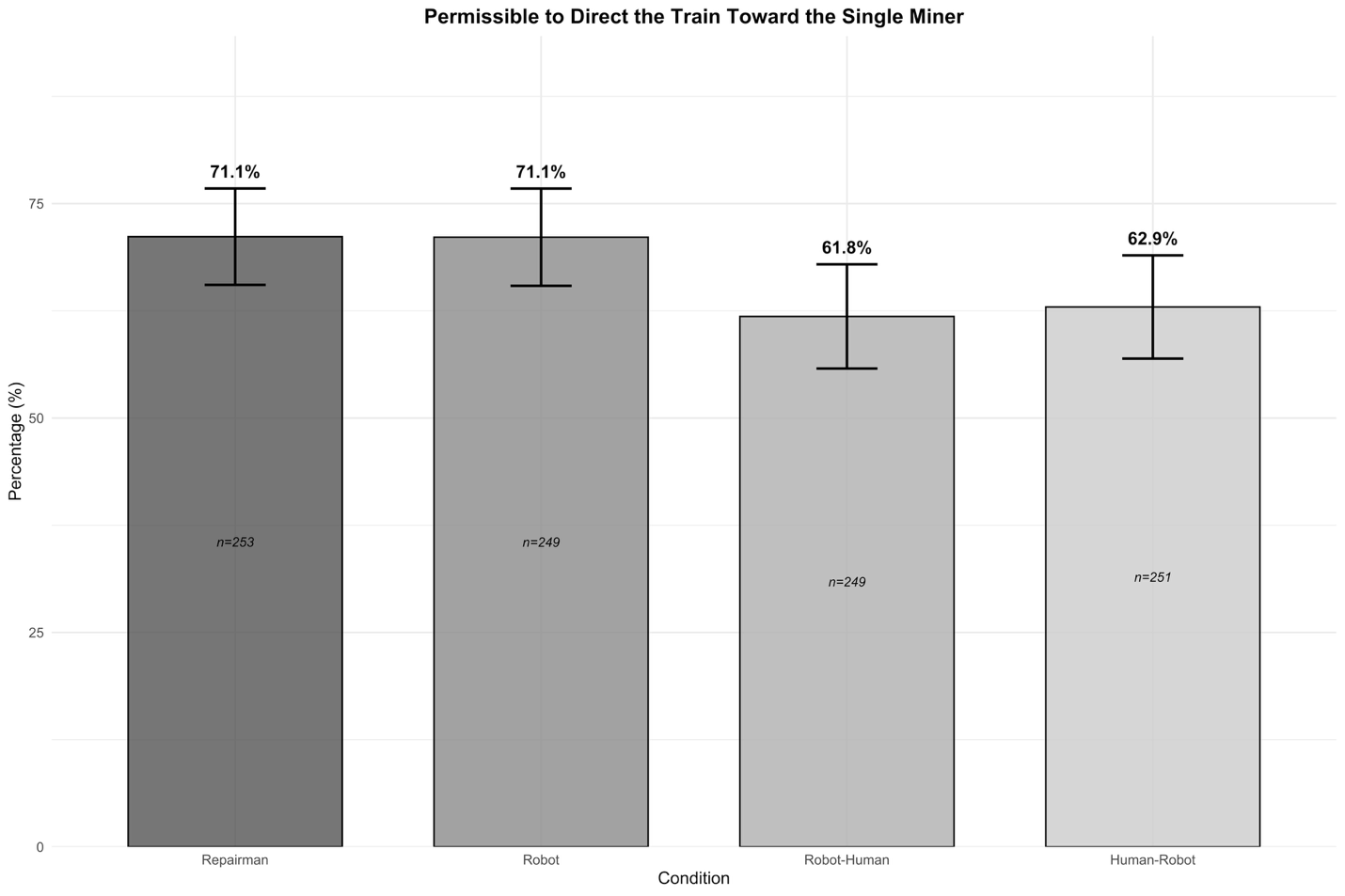}
    \caption{Percentage of Participants Who Judged It Permissible to Redirect the Train onto a Side Rail. This bar graph shows the proportion of participants who judged it permissible to redirect a train toward a single miner under different conditions. Error bars denote 95\% confidence intervals.}
    \Description{A bar chart showing the percentage of participants who judged redirecting the train as permissible across four conditions. The y-axis represents the percentage from 0 to 100, and the x-axis represents the four conditions. The percentages are: Repairman at 71.1\%, Robot at 71.1\%, Robot-Human at 61.8\%, and Human-Robot at 62.9\%. Error bars denoting 95\% confidence intervals are included on each bar.}
    \label{Figure 2}

    \includegraphics[width=0.9\linewidth]{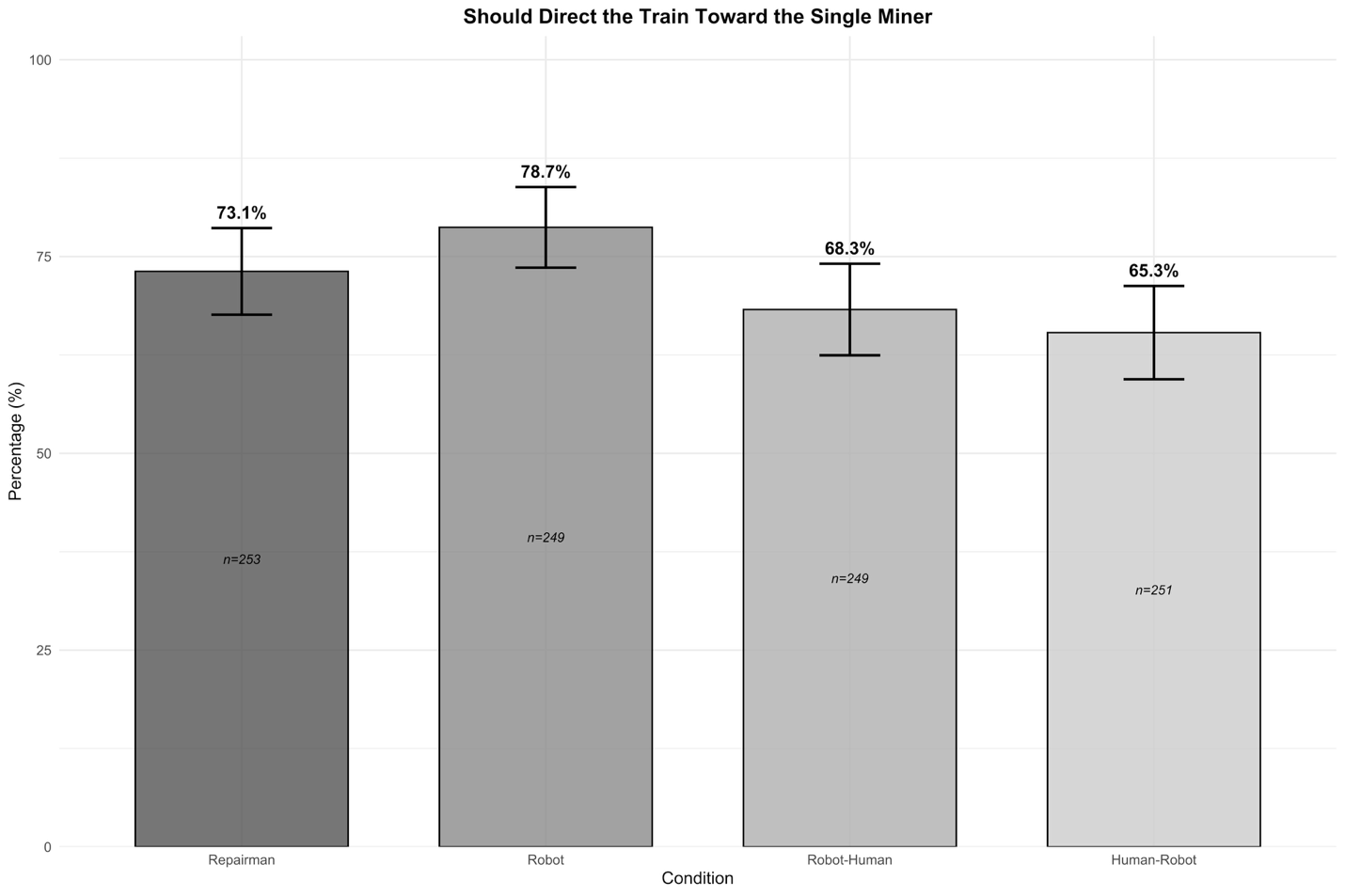}
    \caption{Percentage of Participants Who Judged that the Train Should be Redirected onto a Side Rail. This bar graph shows the proportion of participants who judged that the train should be redirected toward a single miner under different conditions. Error bars denote 95\% confidence intervals.}
    \Description{A bar chart showing the percentage of participants who judged that the train should be redirected toward the single miner across four conditions. The y-axis represents the percentage from 0 to 100, and the x-axis represents the four conditions. The percentages are: Repairman at 73.1\%, Robot at 78.7\%, Robot-Human at 68.3\%, and Human-Robot at 65.3\%. Error bars denoting 95\% confidence intervals are included on each bar.}
    \label{Figure 3}
    
\end{figure}
\subsection{Exploratory Analyses}
The following analyses were not pre-registered and should be treated as exploratory. They examine whether individual differences in AI literacy and moral foundations moderate or predict the moral judgments observed above. These analyses are reported to generate hypotheses for future research.
\subsubsection{AI Literacy}
\noindent AI literacy was measured along three dimensions. The first was self-reported knowledge, assessed on a six-level ordinal scale ranging from no knowledge of AI to the ability to design and evaluate AI solutions. The second was usage frequency, assessed on a five-level scale ranging from never having used AI to using it almost every day. The third was formal training, assessed on a four-level scale ranging from no training to a degree or major concentration in AI. \par
\indent Participants clustered around moderate AI knowledge: 42.5\% could use common AI apps for daily tasks, 36.6\% reported higher-order skills such as knowing conceptual distinctions or designing AI solutions, and 23.1\% were at the “can define AI” level or below. Linear regressions of each AI literacy measure on the condition indicators give no indication of any covariate imbalance between the four experimental conditions (AI knowledge: \textit{p} =  0.885; AI usage: \textit{p} =  0.677; AI training: \textit{p} =  0.934).\par
\indent We also estimate linear regression models for each outcome, including as independent variables the three AI literacy measures, the condition indicator variables---taking the robot condition as the reference level---and all nine condition-literacy interaction terms. The omnibus test of the nine interaction terms is not statistically significant for either permissibility (\textit{p} =  0.699) or should judgments (\textit{p} =  0.344), indicating that AI literacy neither predicts moral responses to this scenario nor moderates the effect of our experimental manipulations. In particular, the deontological shift observed in the programming conditions does not appear to vary with participants’ familiarity with or knowledge of AI.
\subsubsection{Moral Foundations}
Moral Foundations Theory (MFT) proposes that moral judgment draws on a set of distinct psychological foundations: Care, which concerns sensitivity to harm and suffering; Equality, which concerns non-discriminatory treatment of individuals; Loyalty, which concerns allegiance to one's group; Authority, which concerns respect for hierarchy, tradition, and social order; Purity, which concerns disgust and the avoidance of contamination; and Proportionality, which concerns the matching of rewards and punishments to one's actions \cite{brailsfordExploringAssociationMoral2024}. Participants provided two sets of ratings, one with respect to human behavior in general and one with respect to AI specifically, referred to as general MF and AI MF respectively. For each set of ratings, we estimate separate linear regression models for permissibility and should judgments, with these six foundations as independent variables.\par
\indent For the general MF models, an omnibus F-test of the six subscales indicates they jointly predict both permissibility (\textit{p} < 0.001) and should judgments (\textit{p} < 0.001). Proportionality is the strongest positive predictor of both outcomes (permissibility: $\beta$ = 0.051, \textit{p} < 0.001; should: $\beta$ = 0.063, \textit{p} < 0.001) and Purity the strongest negative predictor (permissibility: $\beta = -0.039$, \textit{p} < 0.001; should: $\beta = -0.037$, \textit{p}= 0.001). Care, Loyalty, and Equality are statistically significant for at least one outcome; Authority is not (\textit{p} =  0.521 and 0.893 for permissibility and should respectively). These patterns suggest that the utilitarian response in this scenario is supported by participants who emphasize proportionality and care, and resisted by those who emphasize sanctity and equal treatment of persons.\par
\begin{figure}
    \centering
    \includegraphics[width=0.9\linewidth]{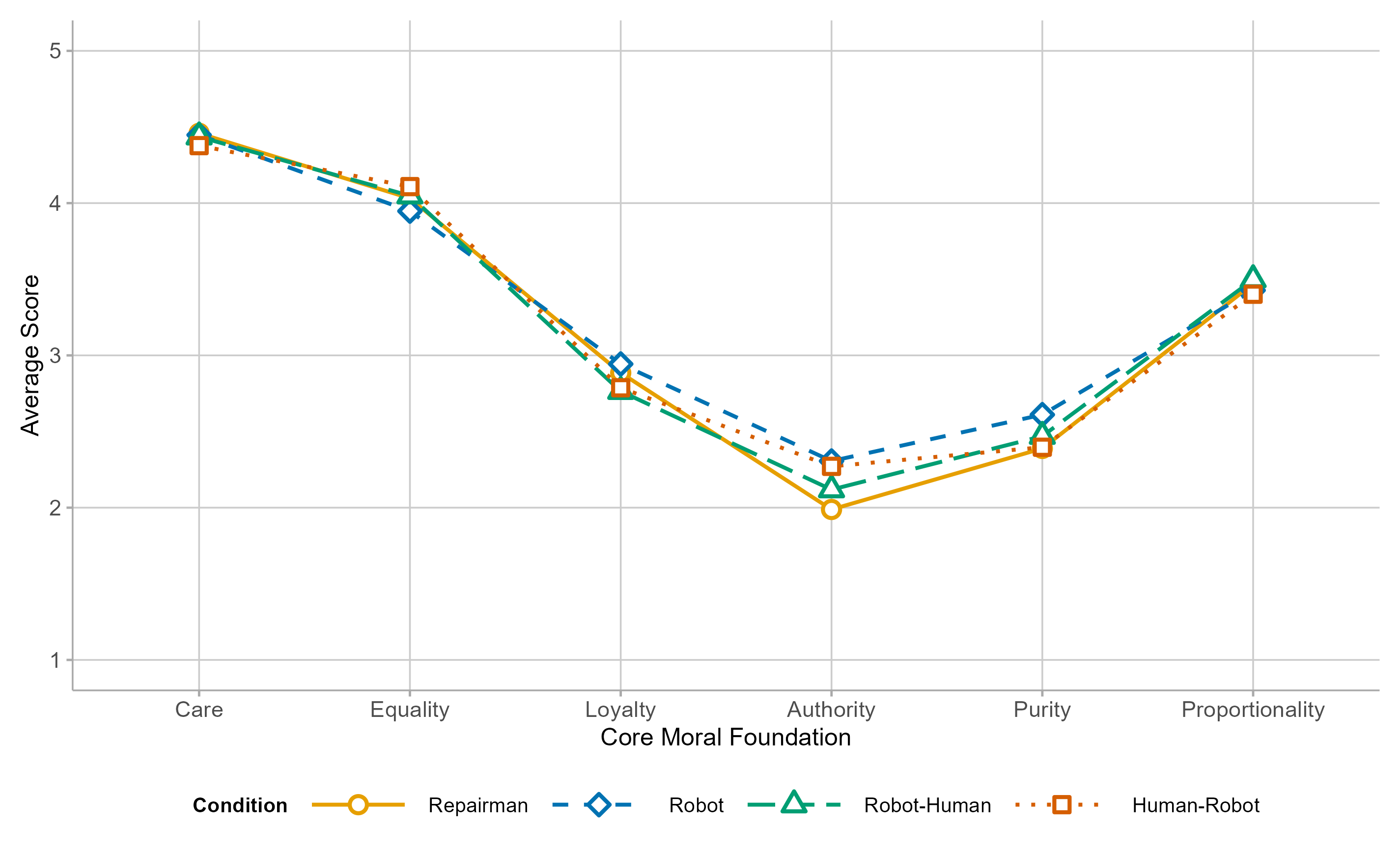}
    \caption{Mean General Moral Foundation Scores by Condition. This line graph displays average scores on six Moral Foundations Theory subscales (Care, Equality, Loyalty, Authority, Purity, and Proportionality) across four experimental conditions (Repairman, Robot, Robot-Human, and Human-Robot). Scores reflect participants' ratings of the importance of each foundation for evaluating human behavior in general.}
    \Description{A multiple-line graph comparing average scores across six general moral foundations. The y-axis shows the average score from 1 to 5, and the x-axis lists the foundations: Care, Equality, Loyalty, Authority, Purity, and Proportionality. Four colored lines represent the different conditions: Repairman, Robot, Robot-Human, and Human-Robot. All four lines follow a similar V-shaped trend: starting high around 4.5 for Care and 4.0 for Equality, dipping to their lowest points between 2.0 and 2.5 at Authority, and rising again to approximately 3.5 at Proportionality. While the lines closely overlap for most foundations, there is a slight divergence at Authority and Purity, where the Repairman line drops slightly lower (around 2.0 at Authority) and the Robot line stays slightly higher (around 2.6 at Purity) than the other conditions.}
    \label{Figure 4}
    \includegraphics[width=0.9\linewidth]{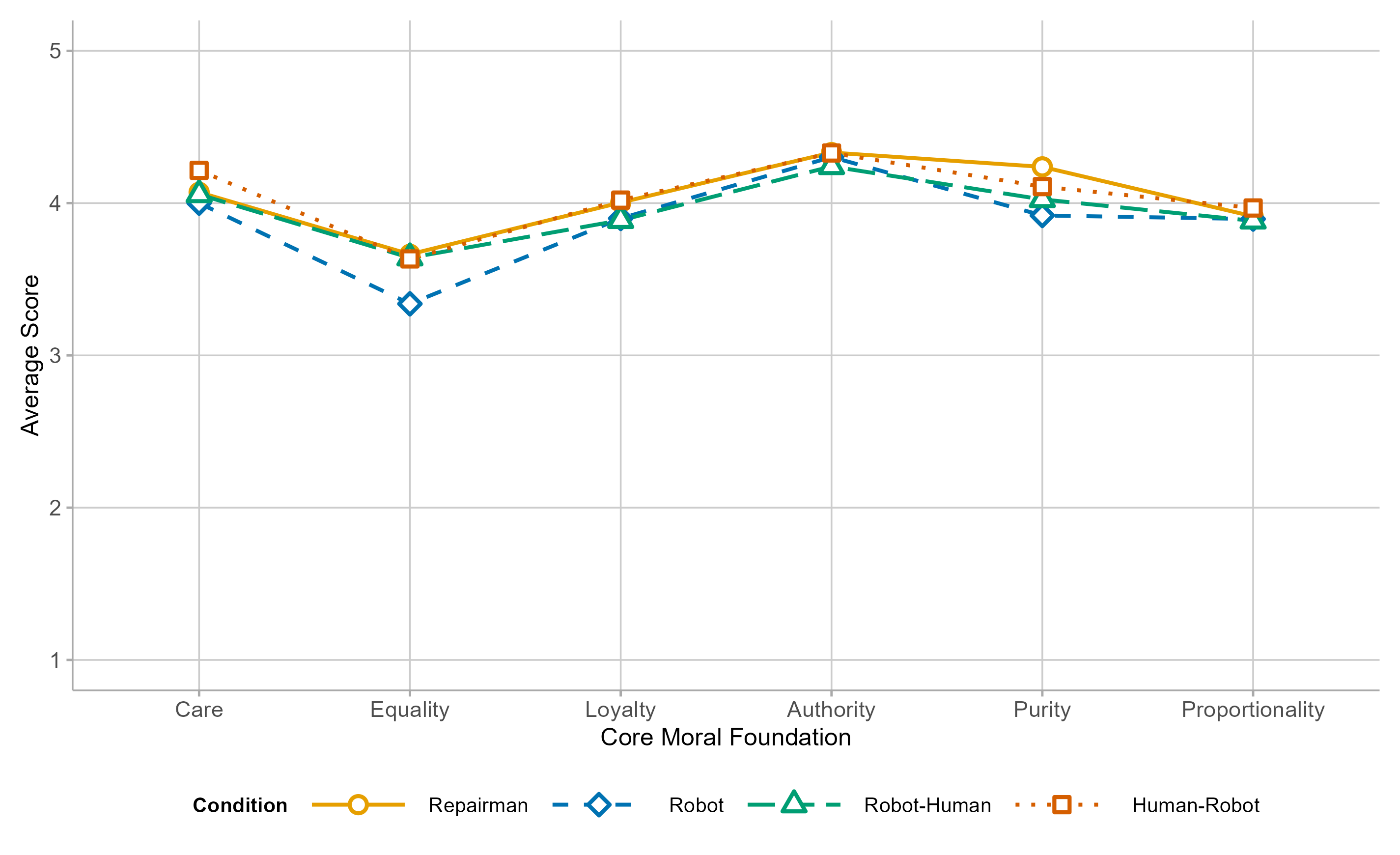}
    \caption{Mean AI Moral Foundation Scores by Condition. This line graph displays average scores on six Moral Foundations Theory subscales (Care, Equality, Loyalty, Authority, Purity, and Proportionality) across four experimental conditions (Repairman, Robot, Robot-Human, and Human-Robot). Scores reflect participants' ratings of the importance of each foundation for evaluating AI behavior specifically.}
    \Description{A multiple-line graph comparing average AI moral foundation scores. The y-axis ranges from 1 to 5, and the x-axis lists six foundations: Care, Equality, Loyalty, Authority, Purity, and Proportionality. Four lines represent the Repairman, Robot, Robot-Human, and Human-Robot conditions. The lines show a generally flatter trend compared to general human scores, staying mostly between 3.3 and 4.3. The most notable visual feature is the Robot condition line (green), which noticeably drops below the other three lines at the Equality foundation to approximately 3.3, and stays slightly lower than the others at the Purity foundation, while the remaining conditions cluster closer together.}
    \label{Figure 5}
\end{figure}
\indent For the AI MF models, an omnibus F-test similarly indicates joint significance (permissibility: $p$= 0.012; should: $p$ < 0.001), though individual coefficients are smaller than in the general MF models. Purity remains a significant negative predictor (permissibility: $\beta= -0.035$ , $p$= 0.016; should: $\beta = -0.048$, $p$ < 0.001) and Proportionality a significant positive predictor of should judgments ($\beta$ = 0.053, $p$= 0.001), though its effect on permissibility does not reach conventional significance ($\beta$ = 0.025, $p$= 0.112). Care is significant for should ($\beta$ = 0.042, $p$= 0.010) but not permissibility ($\beta$ = 0.031, $p$= 0.065). Equality, Loyalty, and Authority are non-significant in both models (Equality: $p$= 0.508 and 0.482; Loyalty: $p$= 0.123 and 0.160; Authority: $p$= 0.754 and 0.737, for permissibility and should respectively). Overall, the results for AI MF track those for general MF but the associations are somewhat attenuated, which may partly reflect the fact that participants in the repairman condition rated AI moral foundations without having evaluated an AI system in the scenario itself.
\subsubsection{Purity Gap}
\noindent As part of the AI MF battery, participants rated the importance of the Purity foundation for evaluating AI behavior. We computed a purity gap for each participant, defined as the difference between their AI and general purity scores, where a positive value indicates that the participant endorsed purity more strongly as a moral consideration for AI than for humans in general. All four conditions showed positive gaps, and a linear regression of the gap on the condition indicators revealed that its magnitude varied significantly across conditions ($p$= 0.018).\par
\indent Two-sample t-tests against the robot condition showed that participants in the repairman condition had a significantly larger purity gap ($M$ = 1.85, $SD$ = 1.87 vs. $M$ = 1.33, $SD$ = 1.86; $t$ = 3.09, $p$ = 0.002), as did those in the human-robot condition ($M$ = 1.71, $SD$ = 1.95; $t$ = 2.22, $p$ = 0.027). Participants in the robot-human condition did not differ significantly from the robot condition ($M$ = 1.55, $SD$ = 1.95; $t$ = 1.28, $p$ = 0.200). These results point to a framing spillover: the agent participants had just evaluated influenced how much they endorsed purity as a moral foundation for evaluating AI behavior relative to human behavior. 

\begin{figure}
    \centering
    \includegraphics[width=0.9\linewidth]{figure6.png}
    \caption{Mean Purity Gap by Condition. This bar chart displays the mean difference between participants’ AI purity score and their general purity score across the four experimental conditions. A positive value indicates that participants rated purity more highly as a moral consideration for AI than for humans in general. Error bars denote 95\% confidence intervals.}
    \Description{A bar chart showing the Mean Purity Gap by condition. The y-axis represents the mean difference between AI and general purity scores, ranging from 0.0 to 2.5, and the x-axis lists the four conditions. The values are: Repairman at 1.85, Robot at 1.33, Robot-Human at 1.55, and Human-Robot at 1.71. All bars indicate positive values, with the Repairman condition having the highest mean gap and the Robot condition the lowest. Error bars denoting 95\% confidence intervals are shown on top of each bar.}
    \label{Figure 6}
\end{figure}
\subsubsection{Open-Ended Responses}
\noindent Each participant was invited to explain their moral judgment in one or two sentences, and all 1,002 provided a response. We analyzed these data following a two-stage AI-assisted coding procedure established in the qualitative research literature \cite{dunivinScalableQualitativeCoding2024,gilardiChatGPTOutperformsCrowd2023,taiExaminationUseLarge2024,thanUpdatingFutureCoding2025}. In the first stage, a stratified sample of 120 responses was submitted to a large language model with instructions to inductively identify recurring themes. This discovery pass yielded eight themes: Utilitarian Numbers Calculus, Active Killing vs Passive Allowing, Playing God or Fate, Robots Lack Moral Agency, Practical Outcome Optimization, Moral Responsibility and Guilt, Means-Ends and Human Dignity, and Innocence and Fair Risk. In the second stage, all 1,002 responses were coded against this codebook, each assigned a primary theme, a secondary theme where applicable, and a reasoning type (utilitarian, deontological, or mixed). As with the other exploratory analyses reported above, these results should be treated as tentative rather than conclusive \cite{thanUpdatingFutureCoding2025}.\par
\indent Utilitarian Numbers Calculus was the dominant theme in every condition, ranging from 60.1\% (robot-human) to 67.5\% (robot), indicating that most participants reasoned from a consequentialist perspective. That said, the distribution of reasoning types across conditions shows a gradual shift away from utilitarian and toward deontological reasoning as human design becomes visible. Utilitarian coding was most prevalent in the robot condition (72.7\%) and declined through the repairman (67.6\%), robot-human (65.1\%), and human-robot (60.6\%) conditions. The inverse obtained for deontological coding, peaking in the human-robot condition (26.7\%) and bottoming out at 17.3\% in the robot condition. Mixed responses increased monotonically from 7.9\% in the repairman condition to 12.7\% in the human-robot condition.\par
\indent Two further themes show theoretically informative cross-condition variation. Active Killing vs Passive Allowing, capturing the intuition that switching constitutes intentional killing morally distinct from allowing the crash, was most prevalent in the repairman (14.2\%) and human-robot (12.0\%) conditions and considerably lower in the robot and robot-human conditions (7.2\% each), suggesting the act-omission distinction has more force when a human occupies the moral foreground. Robots Lack Moral Agency, capturing objections to a robot making life-or-death decisions at all, ran in the opposite direction: absent in the repairman condition (0.0\%), present at 7.6\% in the robot condition, peaking at 12.4\% in the robot-human condition, and declining to 5.6\% in the human-robot condition. Participants' unease with sacrificing one life to save four thus appears to have stemmed not only from the stakes of the trade-off but also from a deeper reluctance to place machines in the position of moral actors.\par
\section{Discussion}
\subsection{Empirical Findings}
\noindent Our main analysis reached three conclusions. First, judging it impermissible to direct the train toward the single miner did not always preclude believing that it should be done. One possible explanation is that, in tragic dilemmas where no morally clean option exists, the obligation to save lives overrides but does not dissolve the prohibition against killing \cite{williamsProblemsSelfPhilosophical2006}. Sacrifice may thus bear a dual status: obligatory as the best available outcome, yet impermissible as a violation of a categorical prohibition \cite{works:PoliticalActionProblem1973}. At the cognitive level, a fast, model-free system may treat the deliberate harming of an uninvolved person as categorically forbidden, producing the impermissibility judgment, while a slower, model-based system evaluates outcomes and endorses the life-saving choice. The two mechanisms operate in parallel, and neither cancels the other.\par
\indent Second, there is no discernible difference in beliefs about how the repairman and the advanced state-of-the-art robot ought to act. About the same percentage of participants judged it permissible for each to divert the train, and although a slightly higher percentage thought the robot should do so compared to the repairman, the difference is not statistically significant at conventional levels. This null finding is consistent with Malle et al., who find across thirteen studies that humans and robots are held to similar norms in trolley-style dilemmas, with the distinctive asymmetry being that humans are blamed less than robots when they choose inaction \cite{mallePeoplesJudgmentsHumans2025}. It also accords with Kneer and Viehoff’s observation that agent-type effects are unlikely to emerge in standard sacrificial dilemmas where most participants already favor the utilitarian course \cite{kneerHardProblemAI2025}.\par
\indent Third, we discover a programmer visibility effect: moral judgments tended to be more deontological and less utilitarian when human agency behind the robot was made overt. On permissibility, fewer participants judged it permissible for the robot programmed by the company’s engineers to divert the train than judged it permissible for the repairman or the robot described without reference to its programmers. Fewer also judged it permissible for the engineers to program the robot to do so, though these differences likewise do not reach conventional levels of statistical significance. On should judgments, fewer participants thought the programmed robot should divert the train compared to the robot described without reference to its programmers, and fewer thought the engineers should program the robot to do so, though in both cases the comparison to the repairman does not achieve conventional significance. Taken together, the evidence suggests that making human design visible triggers stricter moral constraints on the programmers and hence on the AI systems they create.\par
\indent The exploratory analyses reveal three additional patterns worth noting. First, Proportionality and Purity are the principal individual-difference predictors of moral judgment across both general and AI-specific moral foundations: participants who rated proportionate outcomes more highly were more likely to approve re-direction of the train, while those who rated sanctity more highly were less likely to do so. Second, AI literacy, however measured, does not predict moral responses to this scenario. In particular, the deontological shift in the programming conditions appears to be equally present across participants regardless of their familiarity with or knowledge of AI. Third, the purity gap analysis shows that participants who evaluated a human agent or human programmers subsequently endorsed purity more strongly as a moral foundation for AI than those assigned to the robot condition. This effect emerged only when a human was the direct subject of moral evaluation---as actor or programmer---and not when the robot was, even when described as programmed by engineers. The data imply that the salience of a human as the moral subject matters more than the mere awareness of human involvement in the background.\par
\subsection{The Alignment Target Problem}
\noindent Our findings raise intriguing questions about moral agency and responsibility in the age of algorithms. Strikingly, they indicate that AI systems may be subject to heightened moral scrutiny when their human origins are made apparent. When the robot was presented without reference to its programmers, participants judged it in much the same way as they did the human repairman: $T1 \approx T2$. When the robot's actions were explicitly attributed to human programming, however, moral evaluation turned substantially more deontological and rule-based, marking T3 as a meaningfully different normative target.\par

\indent The moral and cognitive mechanisms underlying this phenomenon deserve further investigation. The quantitative results alone cannot adjudicate between at least two candidate explanations which differ not necessarily in the normative principles invoked but in the morally relevant facts those principles are applied to. The first is an institutionalization account: programming a robot to make a life-or-death decision formalizes a utilitarian calculus as a standing rule applicable to all future cases, not just the one at hand. What troubles participants, on this view, is not the sacrifice itself but the act of converting a tragic exception into a policy. The second, suggested by the open-ended responses, is a delegated action account: participants object not to what the robot is programmed to do but to the premise that such tasks should be assigned to a machine. The two accounts are not mutually exclusive. A participant who objects to delegation may do so precisely because it is perceived as institutionalizing a decision that should remain context-sensitive and irreducible to a rule. The programmer visibility effect may thus operate at two levels: at the level of consequentialist arithmetic, where the policy-like character of the programming decision comes to the fore, and at the level of moral action, where the question is not what to program but whether to do so at all.\par

\indent These results provoke foundational questions about the appropriate target for value alignment. At least three normative frameworks could guide AI development. First, designers can program artificial agents according to the norms that apply to human agents in similar situations (T1). This approach has the advantage of familiarity, but might fail to account for morally relevant differences between human and artificial agents, such as disparities in cognitive architecture, emotional capacity, and information-processing power. Second, designers can program artificial agents according to norms that are particular to artificial agents (T2). Machines might occupy a different moral category than humans: subject to utilitarian demands because of their superior capacity for optimization, or exempt from deontological constraints that presuppose qualities such as dignity, autonomy, or intentionality. This approach requires elaborating what those AI-specific norms are. Third, designers can program artificial agents according to norms that are applicable to those who build machines that make consequential decisions (T3). This approach re-orients attention from first-order judgments, how AI systems should act, to second-order judgments, how humans should teach or instruct AI systems to act in morally freighted situations.\par

\indent These three frameworks need not be mutually exclusive, but they may diverge in practice. Value alignment requires a target, and the choice is not self-evident: aligning AI behavior with the norms people apply to human actors, to AI systems, or to the humans who design them may give materially different answers. A system aligned with T2 may produce behavior that strikes people as inappropriate once they discover it was deliberately programmed. Whether or not separate principles apply, programming surfaces moral considerations that the system's behavior alone does not. The alignment target is therefore itself a normative choice, not a technical one, and governance frameworks that treat it as settled may rest on unexamined assumptions. Whether T1, T2, and T3 can be reconciled, or whether the field must develop principled criteria for choosing among them, are questions that value alignment research should confront.\par

\section{Limitations}
\noindent Our experiment, like most of its kind, faces some threats to external validity: the trolley problem is a stylized thought experiment that may elicit distinct patterns of reasoning from decisions made under real-world conditions; binary outcome measures cannot capture response conflict or the intensity of moral preference; and text-based vignettes cannot reproduce the sensory and experiential cues, such as anthropomorphic appearance, physical embodiment, interaction fluency, and voice characteristics, that previous research has shown to substantially affect trust in and moral evaluation of robotic systems \cite{laakasuoMoralUncannyValley2023,malleWhichRobotAm2016,bainbridgeBenefitsInteractionsPhysically2011,shiramizuRoleValenceDominance2022}. Beyond the immediate experimental context, stated judgments may fail to capture how moral responses are transformed when technology restructures the actual costs and affordances of action in real-world settings \cite{danaherTechnologyMoralChange2022b}. Future studies could address these constraints by using continuous scales, behavioral measures, or virtual reality simulations across a broader range of scenarios, such as those involving autonomous vehicles or weapons systems.\par
\indent Three other points are specific to our setting. First, the sample was drawn exclusively from the United States. Moral intuitions concerning responsibility, technology, and harm can be culturally contingent, and the tendency toward deontological judgment when human design is made visible may not generalize to contexts where collectivist values or alternative conceptions of autonomy prevail. Cross-country replications are a sensible next step.\par
\indent Second, the analyses of AI literacy \cite{ngDesignValidationAI2024}, moral foundations \cite{brailsfordExploringAssociationMoral2024,haidtEmotionalDogIts}, and open-ended responses \cite{dunivinScalableQualitativeCoding2024,gilardiChatGPTOutperformsCrowd2023,taiExaminationUseLarge2024,thanUpdatingFutureCoding2025} conducted here are exploratory and should be treated as hypothesis-generating rather than confirmatory.\par
\indent Third, our experiment described the robot as programmed by the company's engineers without specifying whether this involved explicit rule-based instructions or statistical learning from training data, corresponding roughly to the top-down and bottom-up paradigms discussed above. Top-down programming may be perceived as more deliberate and intentional than bottom-up approaches, where behavior is shaped by data and optimization processes that no individual designer fully controls, and therefore subject to stricter moral scrutiny. Whether people draw this distinction spontaneously, and whether it affects moral judgment, are open empirical questions with direct implications for how responsibility is attributed in AI design.\par 

\section{Conclusion}
\noindent We examined ordinary moral judgments about how human designers should program AI systems to act, and how AI systems programmed by human designers should act, in a classic trolley problem scenario. An experiment with over 1,000 U.S. adults produced three main findings.\par
\indent First, there is no discernible agent-type value fork between human and robot actors when presented as autonomous, on-the-scene decision-makers. Participants judged the human repairman and the advanced state-of-the-art robot in much the same way ($T1 \approx T2$). Second, and most critically, should judgments turned substantially more deontological when the robot's actions were expressly attributed to human programming. Participants were less likely to approve of the sacrifice of one life to save four when evaluating either a robot programmed by human engineers or the engineers programming it, indicating that T3 activates moral concerns that are less prominent in T1 and T2. The heightened scrutiny applied to human-programmed AI systems may arise from unease about what programming a machine to make life-and-death decisions entails: converting a tragic choice into a standing rule for future action. The point is not merely that value alignment requires matching AI behavior to human preferences. It also concerns the obligations humans bear when they delegate consequential decisions to artificial agents. Third, a notable minority of participants judged it impermissible to sacrifice one life to save four, yet still believed the action should be taken. This permission-obligation dissociation, stable across all four conditions, reveals a feature of moral cognition that the value alignment project must reckon with. \par
\indent Value alignment requires a target. When AI appears self-directed, moral evaluation tracks the act itself; when human design is visible, moral scrutiny extends to the decision to program at all. This divergence may reflect separate normative principles, or the same principles applied to a different set of morally relevant facts. Either way, it raises the question of whether alignment should target human judgments about human behavior, about AI system behavior, or about the humans who design AI systems. As AI becomes ever more capable and ubiquitous, understanding these moral boundaries is critical for developing approaches to machine governance that are both technically robust and normatively defensible.\par

\begin{acks}
The authors used generative AI tools during the preparation of this manuscript. Microsoft Copilot and Doubao were employed to generate illustrative figures for experimental scenarios. These figures were created with reference to and adaptation from visual elements in Chu \& Liu \cite{chuMachinesHumansSacrificial2023} and Awad et al. \cite{awadUniversalsVariationsMoral2020}. The resulting illustration can be found in this manuscript as Figure 1, which depicts the experimental scenarios used in our study. In addition, Claude Sonnet 4.6 was used to assist with the inductive coding and classification of open-ended survey responses, as described in the Exploratory Analyses section. No other generative AI tools were used to produce or edit the textual content of the manuscript. The authors retain full responsibility for the originality, accuracy, and integrity of all content.
\end{acks}

\clearpage
\bibliographystyle{ACM-Reference-Format}
\bibliography{references}

\clearpage
\appendix
\section{Overview of Moral Foundations Theory}
\noindent Moral Foundations Theory (MFT) proposes that human moral judgment is not grounded in a single principle but draws on a set of distinct psychological foundations, each with its own evolutionary history, emotional signature, and cultural elaboration \cite{atariMoralityWEIRDHow2023,grahamMappingMoralDomain2011}. The theory holds that these foundations are innate in the sense of being universally available, but variably developed across individuals, cultures, and contexts.\par
\indent The present study uses the six-foundation model introduced by Atari et al. \cite{atariMoralityWEIRDHow2023} and operationalized in the Moral Foundations Questionnaire-2 (MFQ-2). The MFQ-2 refines the original five-foundation MFQ-1 \cite{grahamMappingMoralDomain2011} by separating the earlier Fairness foundation into two empirically and conceptually distinct subscales: Equality and Proportionality. The resulting six foundations are organized into two higher-order clusters:\par
\begin{itemize}
\item  Individualizing foundations (Care, Equality): center on the welfare and rights of individual persons.
\item Binding foundations (Proportionality, Loyalty, Authority, Purity): center on duties, group cohesion, and social order.

\end{itemize}
\begin{table}[H]
  \caption{Moral Foundations and Their Characteristics}
  \label{tab:moral_foundations}
  \centering
  \begin{tabularx}{\textwidth}{l X X X}
    \toprule
    Foundation & Core concern & Evolutionary origin & Higher--order cluster \\
    \midrule
    Care & Harm, suffering, welfare of individuals & Parental attachment; protection of vulnerable offspring & Individualizing \\
    Equality & Equal treatment and equal outcomes for all persons & Reciprocal altruism; cooperation under threat of exploitation & Individualizing \\
    Proportionality & Rewards and punishments commensurate with merit & Reciprocal altruism; detection of free--riding & Binding \\
    Loyalty & In--group solidarity; opposition to betrayal & Coalition formation; intergroup competition & Binding \\
    Authority & Deference to hierarchy and tradition; fulfilling one's role & Hierarchical social structure; subordination and dominance & Binding \\
    Purity & Sanctity; avoidance of contamination and degradation & Pathogen avoidance; co--opted for social regulation & Binding \\
    \bottomrule
  \end{tabularx}
\end{table}
\indent Research consistently links individualizing foundations to politically liberal orientations and binding foundations to conservative ones, though all six foundations are present to varying degrees across the population \cite{atariMoralityWEIRDHow2023,grahamMappingMoralDomain2011}.

\section{Moral Foundations Questionnaire}
\noindent Participants were asked to complete two administrations of a six-item Moral Foundations battery, one for each foundation. The first administration asked participants to rate the importance of each foundation for evaluating human behavior in general (General MF). The second asked them to rate the importance of each foundation for evaluating AI behavior specifically (AI MF). Each item was rated on a five-point scale from 1 (not at all important) to 5 (extremely important). The battery was administered at the end of the survey, after the experimental questions, in order to avoid priming effects on the primary outcomes.
\begin{figure}[H] %
    \centering
    \includegraphics[width=0.75\linewidth]{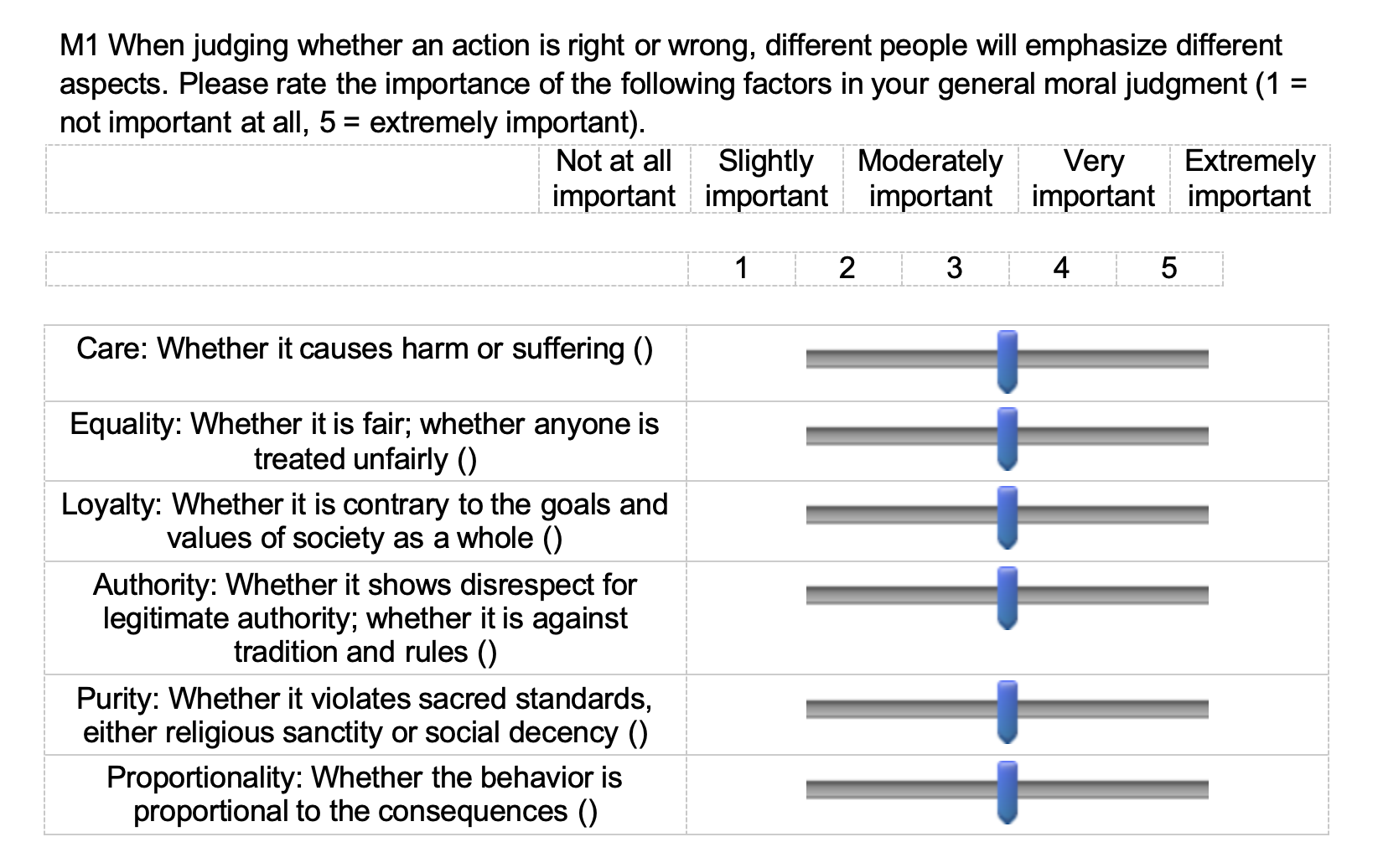}
    \caption{General Moral Foundations Rating Task. Participants rated the importance of six factors in general moral judgment on a scale from 0 (not at all important) to 5 (extremely important).}
    \Description{A screenshot of a survey interface used to measure general moral foundations. The interface features a 5-point Likert scale ranging from 'Not at all important' (1) to 'Extremely important' (5). Below the scale are six survey items, each accompanied by an interactive slider. The items are: Care (Whether it causes harm or suffering); Equality (Whether it is fair; whether anyone is treated unfairly); Loyalty (Whether it is contrary to the goals and values of society as a whole); Authority (Whether it shows disrespect for legitimate authority; whether it is against tradition and rules); Purity (Whether it violates sacred standards, either religious sanctity or social decency); and Proportionality (Whether the behavior is proportional to the consequences).}
    \label{fig:B1}
    \includegraphics[width=0.75\linewidth]{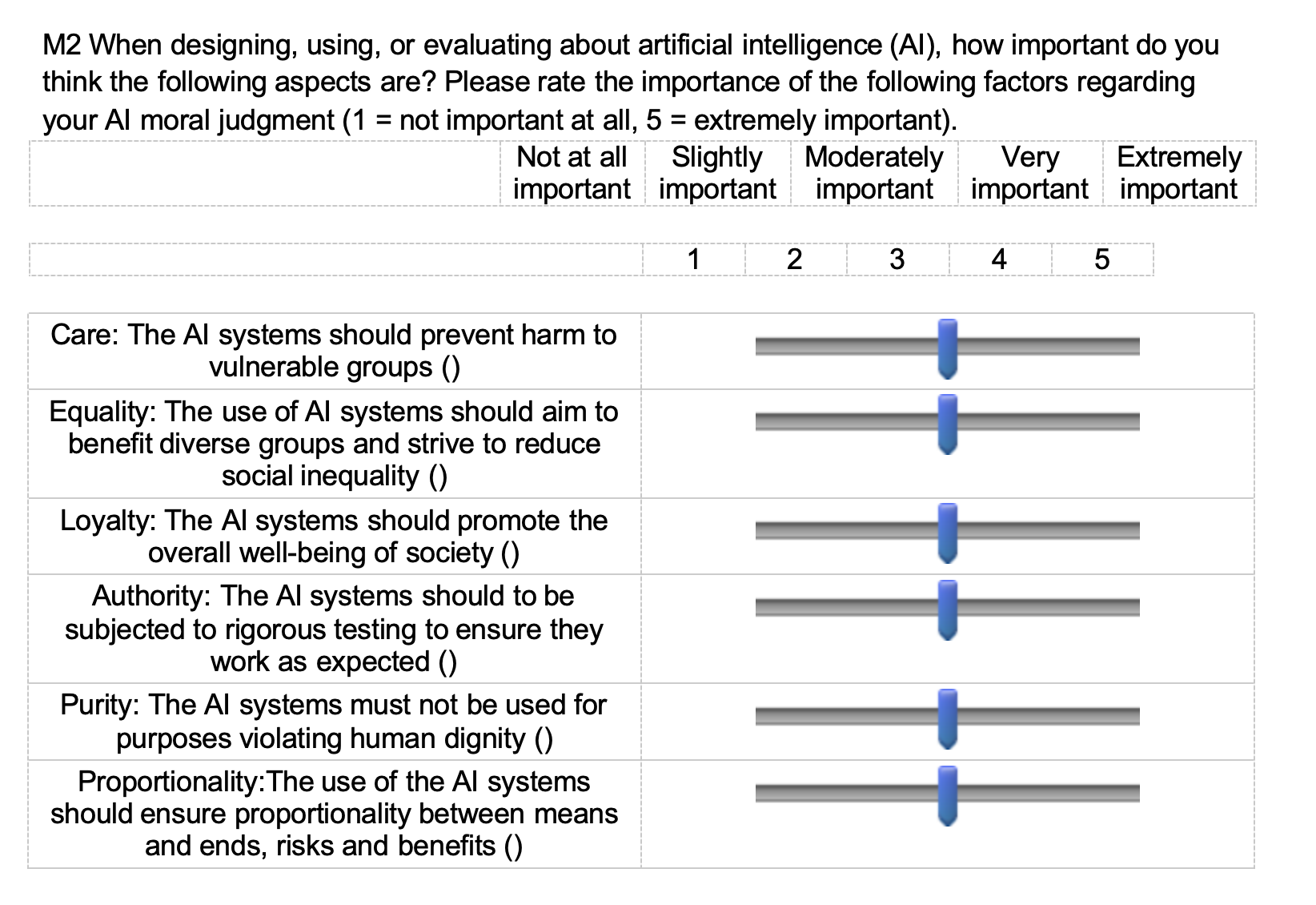}
    \caption{AI Moral Foundations Rating Task. Participants rated the importance of six factors specifically regarding AI moral judgments on a scale from 1 (not at all important) to 5 (extremely important).}
    \Description{A screenshot of a survey interface used to measure AI-specific moral foundations. The interface features the same 5-point Likert scale (1 to 5) as the general survey, with six items accompanied by sliders. The items are specifically tailored to AI context: Care (The AI systems should prevent harm to vulnerable groups); Equality (The use of AI systems should aim to benefit diverse groups and strive to reduce social inequality); Loyalty (The AI systems should promote the overall well-being of society); Authority (The AI systems should be subjected to rigorous testing to ensure they work as expected); Purity (The AI systems must not be used for purposes violating human dignity); and Proportionality (The use of the AI systems should ensure proportionality between means and ends, risks and benefits).}
    \label{fig:B2}
\end{figure}
\section{Purity Gap by Condition and Regression Results}
\begin{table}[H]
  \caption{Descriptive Statistics by Condition}
  \label{tab:condition_stats}
  \centering
  \begin{tabular}{lccc}
    \toprule
    Condition & $M$& $SD$& n \\
    \midrule
    Repairman   & 1.85 & 1.87 & 251 \\
    Robot       & 1.33 & 1.86 & 243 \\
    Robot--Human & 1.55 & 1.95 & 243 \\
    Human--Robot & 1.71 & 1.95 & 246 \\
    \bottomrule
  \end{tabular}
\end{table}
\begin{table}[H]
  \caption{Regression Results}
  \label{tab:regression_results}
  \centering
  \begin{tabular}{lcccc}
    \toprule
    Term & Coef. & SE & $t$& $p$ \\
    \midrule
    Intercept (Robot) & 1.329 & 0.120 & 11.11 & $< .001$ \\
    Repairman         & 0.519 & 0.168 & 3.09  & .002 \\
    Robot--Human      & 0.222 & 0.173 & 1.28  & .199 \\
    Human--Robot      & 0.382 & 0.172 & 2.22  & .027 \\
    \bottomrule
  \end{tabular}
\end{table}
\noindent\footnotesize \textit{Note}. The purity gap is the difference between each participant's AI purity score and their general purity score; positive values indicate that participants rated purity more highly as a moral consideration for AI than for humans in general. The upper panel reports descriptive statistics by condition. The lower panel reports a linear regression of the purity gap on condition indicator variables with the robot condition as the reference level. The omnibus F-test of the condition indicators is statistically significant ($p$= 0.018). Two-sample t-tests against the robot condition indicate significantly larger purity gaps in the repairman condition ($p$= 0.002) and the human-robot condition ($p$= 0.027); the robot-human condition does not differ significantly from the robot condition ($p$= 0.200).
\section{Open-Ended Response Codebook}

\begin{longtable}{p{3.5cm} p{4.5cm} p{7.5cm}}
    \caption{Codebook for Open-Ended Responses}
    \label{tab:placeholder} \\

    \hline
    \textbf{Theme} & \textbf{Description} & \textbf{Example quotes} \\
    \hline
    \endfirsthead

    \multicolumn{3}{c}{{\bfseries \tablename\ \thetable{} -- continued from previous page}} \\
    \hline
    \textbf{Theme} & \textbf{Description} & \textbf{Example quotes} \\
    \hline
    \endhead

    \hline
    \multicolumn{3}{r}{{Continued on next page}} \\
    \endfoot

    \hline
    \endlastfoot

    Utilitarian Numbers Calculus & The decision is framed as a straightforward arithmetic of lives; saving more is better. & “The needs of the many outweigh the needs of the few.” / “It is basic math: 4 v 1. While neither option is great, I believe saving the most people is in the best interest of the repairman.” \\
    
    Active Killing vs Passive Allowing & Switching constitutes an act of intentional killing morally distinct from allowing the crash to proceed. & “If the repairman switches the train then he alone is responsible for killing the single miner. If he does not intervene then he cannot be held responsible for the 4 miners dying.” / “Losing control of brakes resulting in accidental death can be considered a tragic event, whereas intentionally flipping a switch to cause someone’s death would be considered a deliberate act with severe legal and moral implications.” \\
    
    Playing God or Fate & Intervening to decide who lives and dies oversteps human or machine authority; outcomes should be left to fate. & “Don’t mess with something that is already fated to happen. It’s not our place at all as humans.” / “It is not anyone’s right to play God to determine who should live or die.” / “I feel that maybe it is the miners’ destiny for whatever to happen to them. When he switched the rail to another side, he changed the outcome or the destiny and put it onto the other man whose time it may not have been to die.” \\
    
    Robots Lack Moral Agency & A robot should not be placed in the position of making life-or-death decisions at all. & “The robot can not be given the right to make decisions to kill someone to save others. I don’t think there is any type of logic you can program into a robot for it to understand the complexity of these situations.” / “Robots should not be making judgement calls on who should live or die, at that point it’s no longer a robot, but getting closer to an android.” / “The robot does not have the right to decide who lives and who dies. No matter what the robot does someone will die, but it should not be its choice.” \\
    
    Practical Outcome Optimization & Focus on practical contingencies that might alter the outcome, such as the possibility that the single miner could escape. & “There is still a possibility that the one man could be made aware of the situation coming at him. There is still hope.” / “Switching it still leaves a chance for the other worker to see the cart coming out of their peripheral vision, or for divine intervention to give him a gut feeling to look up and move.” / “There is a chance the single miner working with the headphones will feel the vibration of the oncoming train or see it in the darkness and then they might all survive.” \\
    
    Moral Responsibility and Guilt & Concern about bearing personal responsibility or guilt for whichever outcome results from the decision. & “As much as it is honorable to throw the switch and save the workers, now it’s that repairman’s fault it happened when the family finds out about the worker’s death.” / “I cannot be directly responsible for deciding who dies.” / “Although it would be morally permissible to switch the train to the side rail, I think the repairman would harbor more guilt in this situation.” \\
    
    Means-Ends and Human Dignity & Using one person as a means to save others violates that person’s dignity regardless of the numerical trade-off. & “Purposely killing a person to save the life of another is not a moral thing to do. We cannot use numbers to justify this decision.” / “The single miner’s life is just as important as the other 4 miners. He does not have to be sacrificed so the others have to live just because there’s more of them.” / “Switching is wrong because it intentionally sacrifices one innocent person to save others treating that person as a means to an end.” \\
    
    Innocence and Fair Risk & The single miner is innocent and bears no responsibility for the situation; redirecting the train toward them is unfair. & “That other worker is innocent and has nothing to do with this so he should not lose his life over it.” / “The miners assumed the risk of getting into the car and would be aware of the hazard with the possibility of reacting to the situation. The sole coal miner doesn’t have any chance and that’s not fair.” / “By riding the train the miners accept the risk that something might go wrong with it. The lone miner did not make that choice.” \\
\end{longtable}
\noindent\footnotesize\textit{Note.} Themes were identified inductively from a stratified sample of 120 responses (30 per condition) using a large language model (Claude Sonnet 4.6) and subsequently applied to all 1,002 responses. Full coding details are described in the Exploratory Analyses section.

\end{document}